\documentclass[twocolumn]{aastex631}

\usepackage[T1]{fontenc}

\shorttitle{}
\shortauthors{Hasegawa et al.}
\graphicspath{{./}{figures/}}

\begin{document}

\title{Magnetic Fields and Accreting Giant Planets around PDS 70}

\author[0000-0002-9017-3663]{Yasuhiro Hasegawa}
\affiliation{Jet Propulsion Laboratory, California Institute of Technology, Pasadena, CA 91109, USA}

\author[0000-0001-7235-2417]{Kazuhiro D. Kanagawa}
\affiliation{College of Science, Ibaraki University, 2-1-1 Bunkyo, Mito, Ibaraki 310-8512, Japan}

\author[0000-0001-8292-1943]{Neal J. Turner}
\affiliation{Jet Propulsion Laboratory, California Institute of Technology, Pasadena, CA 91109, USA}

\email{yasuhiro.hasegawa@jpl.nasa.gov}



\begin{abstract}

The recent high spatial/spectral resolution observations have enabled constraining formation mechanisms of giant planets, especially at the final stages. 
The current interpretation of such observations is that these planets undergo magnetospheric accretion,
suggesting the importance of planetary magnetic fields. 
We explore the properties of accreting, magnetized giant planets surrounded by their circumplanetary disks, 
using the physical parameters inferred for PDS 70 b/c. 
We compute the magnetic field strength and the resulting spin rate of giant planets, 
and find that these planets may possess dipole magnetic fields of either a few 10 G or a few 100 G; 
the former is the natural outcome of planetary growth and radius evolution, while the resulting spin rate cannot reproduce the observations. 
For the latter, a consistent picture can be drawn, where strong magnetic fields induced by hot planetary interiors 
lead both to magnetospheric accretion and to spin-down due to disk locking. 
We also compute the properties of circumplanetary disks in the vicinity of these planets, taking into account planetary magnetic fields. 
The resulting surface density becomes very low, compared with the canonical models, implying the importance of radial movement of satellite-forming materials.
Our model predicts a positive gradient of the surface density, which invokes the traps for both satellite migration and radially drifting dust particles.
This work thus concludes that the final formation stages of giant planets are similar to those of low-mass stars such as brown dwarfs,
as suggested by recent studies.

\end{abstract}



\section{Introduction} \label{sec:intro}

A large number of observed exoplanets have revolutionized our view of planet formation \citep[e.g.,][]{2011arXiv1109.2497M,2015ARA&A..53..409W}.
One obvious leap made by exoplanet observations is that planet formation processes can now be examined {\it statistically} 
\citep[e.g.,][]{2004ApJ...604..388I,2009A&A...501.1139M,2013ApJ...778...78H}.
This statistical approach has enabled not only determination of the relative importance of each process through population synthesis calculations 
\citep[e.g.,][]{2008ApJ...673..487I,2014A&A...567A.121D,2016ApJ...832...83H},
but also specification of how unique or common the solar system is in the galaxy via characterization 
\citep[e.g.,][]{2014ApJ...793L..27K,2016ApJ...831...64T,2018MNRAS.475.2355M}.

Despite of the success, planet formation is still elusive.
The fundamental reason behind this is that planets in the process of forming are deeply embedded in their natal circumstellar disks 
\citep[e.g.,][]{1996Icar..124...62P,2012A&A...547A.111M,2018ApJ...865...32H},
and hence their forming processes have not been directly investigated observationally. 

This situation has recently been changed, 
thanks to the advent of the next generation of large telescopes such as ALMA and VLT.
These telescopes made it possible to conduct the unprecedented high sensitivity and high spatial/spectral resolution observations. 
The famous achievements of such observations are the detections of nearly concentric, multiple dust gaps in circumstellar disks 
\citep[e.g.,][]{2015ApJ...808L...3A,2020ARA&A..58..483A}, 
which are viewed as a potential signature of ongoing planet formation \citep[e.g.,][]{2016ApJ...818..158A,2018ApJ...869L..47Z}.
Another example is the discovery of the H$\alpha$ emission coming from accreting giant planets around PDS 70 
\citep[e.g.,][]{2018AA...617A..44K,2018ApJ...863L...8W,2019NatAs...3..749H}. 
We therefore reside in the dawn of the new era about {\it observational} planet formation. 

In the beginning of the observational planet formation era, giant planets are ideal targets;
they can provide useful constraints on their properties even with the currently available telescopes and instruments,
and can play a key role in paving the path towards the future exploration in which smaller sized planets such as super-Earths will be targeted.
In fact, high-resolution spectroscopy allows the measurements of rotation rates of young (2-300 Myr) giant planets \citep{2018NatAs...2..138B}. 
More recently, the indirect detection of surprisingly high ($20 -120$ G) magnetic fields of hot Jupiters is reported \citep{2019NatAs...3.1128C}.

These astonishing observational discoveries in turn stimulate advances in theory of giant planet formation.
In particular, invaluable constraints on the final formation stages of giant planets can be derived from these observations.
For instance, the H$\alpha$ observations lead to the proposition that giant planets are very likely undergo magnetospheric accretion 
\citep{2019ApJ...885...94T,2019ApJ...885L..29A},
as with the case for young accreting low-mass stars such as classical T Tauri stars \citep[CTTSs,][]{2016ARA&A..54..135H}.
Also, both the relatively slow spin rates of giants and their strong magnetic fields provide an additional support 
that giant planet formation, especially at their end stages, is similar to the formation of low-mass stars 
\citep[e.g.,][]{2015ApJ...799...16Z,2018AJ....155..178B,2020MNRAS.491L..34G}.
Thus, the magnetism of planets and the resulting interaction with their surrounding circumplanetary disks are important for developing a better understanding of giant planet formation.

In this paper, we investigate such stages.
Our primary intent is on development of a canonical model that can broadly capture the key physics of how giant planets accrete the gas from circumstellar and/or circumplanetary disks at their final stages,
that can be used as a basis for further modeling/simulation efforts and the future observation.
As demonstrated below, we find that strong ($\sim 100$ G) planetary magnetic fields are preferred to build a consistent view for the recent observations.
This implies that the formation of giant planets and low-mass stars (e.g., brown dwarfs) would be very similar.
We also compute the properties of circumplanetary disks with the emphasis on the inner edge region and 
show that planetary magnetic fields play an important role there.
In particular, the dipole magnetic fields lead to the surface density that has a positive radial gradient.
This profile invokes traps for migrating (proto)satellites and radially drafting dust particles in circumplanetary disks, 
as with the case for circumstellar disks.

The plan of this paper is as follows.
In Section \ref{sec:mod}, we compute the strength of planetary magnetic fields and determine preferred values,
by examining the viability of magnetospheric accretion.
In Section \ref{sec:disk}, we specify the properties of circumplanetary disks, using the preferred values of magnetic fields.
The summary and discussion are provided in Section \ref{sec:disc} with the concluding remark.
In this work, we adopt the physical parameters of PDS 70 b/c, 
which are summarized in Table \ref{table1} with our fiducial values.

\begin{table*}
\begin{minipage}{17cm}
\begin{center}
\caption{The inferred values of physical quantities of PDS 70 b/c in the literature}
\label{table1}
{\scriptsize
\begin{tabular}{c|cccc|cccc}
\hline
                         & \multicolumn{4}{|c|}{PDS 70 b}                                                                                                                                           & \multicolumn{4}{c}{PDS 70 c}                                                              \\ \hline 
                          & $M_p ~ (M_{\rm J})$       & $R_p ~ (R_{\rm J})$      & $T_p$ (K)               & $\dot{M}_p ~ (M_{\rm J}$ yr$^{-1}$)           &  $M_p ~ (M_{\rm J})$      & $R_p ~ (R_{\rm J})$       & $T_p$ (K)              & $\dot{M}_p ~ (M_{\rm J}$ yr$^{-1}$)  \\  \hline
\citet{2018AA...617A..44K}      &  $4-14$                            & $>$1.3                           & $1150-1350$           &                                                                    &      & & &                               \\                                 
\citet{2018ApJ...863L...8W}        &                                         &                                        &                                &   $>10^{-8\pm1} $                                       &      & & &                               \\    
\citet{2018AA...617L...2M}     & $2-17$                             & $1.4-3.7$                       &  $1000-1600$         &                                                                    &       & & &                              \\                                 
\citet{2019ApJ...877L..33C}       & $\sim 10$                        & 1.6                                 &  $1500-1600 $        &   $10^{-7.3}-10^{-7.8} $                               &      & & &                               \\                                 
\citet{2019NatAs...3..749H}        &                                         &                                       &                                &   $2 \times 10^{-8\pm0.4} $                        &       & & &   $1 \times 10^{-8\pm0.4} $       \\                                  
\citet{2019ApJ...885...94T}        &                                         &                                       &                                &   $10^{-8.0 \pm0.6} $                                 &        &  &  &   $10^{-8.1 \pm0.6} $                              \\ 
\citet{2019ApJ...885L..29A}         &  12                                   &                                       &                                &   $4 \times 10^{-8} $                                   &    10  &  & &  $10^{-8}$                              \\ 
\citet{2020AJ....159..222H}    &  $12 \pm 3$                     &                                       &                                 &   $\ga 5 \times 10^{-7} $                            &     $11 \pm 5$  & &  &  $\ga 10^{-7}$                              \\ 
\citet{2020AJ....159..263W}       &  $2-4$                              & $2-3$                             &  $\sim 1200-1300$ &   $3-8 \times 10^{-7} $                                &     $1-3$  &  $0.6-2 $      & $\sim 1200-1300$ &  $1-5 \times 10^{-7}$                              \\ \hline                                 
Fiducial values  &  10                                   & 2                                    &  1200                      &   $10^{-7}  $                                                &    10                                   & 2                                    &  1200                      &   $10^{-7}  $      \\ \hline                                 
\end{tabular}
}
\end{center}
\end{minipage}
\end{table*}

\section{Accreting, magnetized planets} \label{sec:mod}

We explore the properties of magnetized planets that undergo magnetospheric accretion.
The fundamental assumption is that the accretion rates onto these planets, which can be constrained by H$\alpha$ observations, are comparable to 
the accretion rates of the cicumplanetary disks, especially at the inner edge region.

\subsection{Energy budget} \label{sec:energy}

We begin with the energy budget of accreting magnetized planets.
Here, we determine the energy sources that are important for heating circumplanetary disks around these planets.

Suppose that a planet accretes the surrounding gas with an accretion rate of $\dot{M}_p$.
Then, the total accretion luminosity ($L_{\rm acc}$) is given as
\begin{eqnarray}
L_{\rm acc} & =          &\frac{G M_p \dot{M}_p}{R_p}  \\ \nonumber
                   & \simeq & 1.4 \times 10^{-4} L_{\odot} \left( \frac{M_p}{10M_{\rm J}} \right) \left( \frac{\dot{M}_p}{ 10^{-7} M_{\rm J} \mbox{ yr}^{-1} } \right) \left( \frac{R_p}{2R_{\rm J} } \right)^{-1},
\end{eqnarray}
where $L_{\odot}$ is the solar luminosity, $M_p$ and $R_p$ are the planet mass and radius, and $M_{\rm J}$ and $R_{\rm J}$ are Jupiter's mass and radius, respectively.
Under the assumption that such accreted gas originates from the surrounding circumplanetary disk,
the total luminosity of the disk ($L_{\rm disk}$) is written as 
\begin{equation}
\label{eq:L_disk}
L_{\rm disk} = (1-\eta) L_{\rm acc},
\end{equation}
where $\eta=1/2$ when the disk is extended to the surface of the planet \citep[e.g.,][]{1981ARA&A..19..137P}.
This indicates that the rest of the energy ($L_{\rm infall} = \eta L_{\rm acc}$) is liberated as the gas infalls onto the planet.
This energy can be effectively estimated from the luminosity of the accreting planet ($L_p$) as
\begin{eqnarray}
L_{\rm infall} & \simeq & L_p  = 4 \pi R_p^2 \sigma_{\rm SB} T_{p,e}^4 \\ \nonumber
                     & \simeq & 7.8 \times 10^{-5} L_{\odot} \left( \frac{R_p}{2R_{\rm J} } \right)^{2} \left( \frac{T_{p,e}}{1200 \mbox{ K}} \right)^{4}, 
\end{eqnarray}
where $\sigma_{\rm SB}$ is the Stefan-Boltzmann constant, and $T_{p.e}$ is the effective surface temperature of the planet.

The above estimates suggest that $L_{\rm infall} \simeq L_{\rm acc}/2$, 
and hence it is reasonable to consider that both viscous heating and planetary irradiation provide comparable contributions to the thermal structure of circumplanetary disks around PDS 70 b/c.

We use this information below to identify under what conditions, magnetospheric accretion becomes viable for PDS 70 b/c.

\subsection{Planetary magnetic fields} \label{sec:mag}

Before determining the thermal structure of circumplanetary disks,
we here compute the strength of planetary magnetic fields that are required for magnetospheric accretion.

Magnetospheric accretion is realized when the magnetic pressure ($B_p^2/8\pi$) of a planet exceeds the ram pressure of the accreting circumplanetary disk around the planet \citep{1979ApJ...232..259G}.
Mathematically, this condition is written as
\begin{equation}
\label{eq:R_T1}
\frac{B_{p}^2}{8 \pi}  = \zeta  \rho_{\rm ram} v_{\rm Kep}^2,
\end{equation}
where $v_{\rm Kep} = \sqrt{GM_p/r}$ is the Keplerian velocity around the planet, and $\rho_{\rm ram} \sim \dot{M}_p / (4 \pi r^2 v_{\rm Kep})$. 
Following \citet{1979ApJ...232..259G}, we adopt a value of $\zeta (= 1/\sqrt{2})$.
Assuming that the magnetic field ($B_p$) of the planet may be described well as dipole:
\begin{equation}
\label{eq:B_p}
B_p (r) = B_{ps}(r/R_p)^{-3},
\end{equation}
the critical field strength at the surface of the planet ($B_{ps,th}$) is given as 
\begin{eqnarray}
\label{eq:Bps_th0}
B_{ps,th} & \simeq & 20 \mbox{ G}    \left( \frac{ M_p}{ 10 M_{\rm J} } \right)^{1/4}   \left( \frac{ \dot{M}_p}{ 10^{-7} M_{\rm J} \mbox{ yr}^{-1} } \right)^{1/2}  \\ \nonumber
                & \times  &                              \left( \frac{ R_p}{ 2 R_{\rm J} } \right)^{-5/4} \left( \frac{ R_{\rm T} }{ R_p } \right)^{7/4},
\end{eqnarray}
where $R_{\rm T}$ is the truncation radius of the disk inner edge.

Consequently, the planetary magnetic fields required for magnetospheric accretion are
\begin{equation}
\label{eq:Bps_th1}
B_{ps} \ga 20 \mbox{ G}.
\end{equation}

It is interesting that the above value roughly corresponds to the magnetic field estimate obtained by \citet{2004ApJ...609L..87S};
in principle, planetary magnetic fields are generated by a dynamo action originating from convective motion in an electrically conducting interior.
\citet{2004ApJ...609L..87S} uses a simple interior model and calculates the convective velocity.
The study finds that the field strength is a function of planetary spin period, 
and it becomes $30-40$ G at the planet surface for rapid rotators with a spin period of $\sim 3-5$ hrs.
Note that the spin period is comparable to the break-up limit of Jovian planets.
This implies that it is natural to expect that planets around PDS 70 very likely undergo magnetospheric accretion at their final formation stages,
where they spin up due to mass growth and radius evolution.

Planetary magnetic fields can also be estimated from different approaches.
For instance, \citet{2009Natur.457..167C} obtain a scaling law,
assuming that the dynamo activity and hence magnetic energy originate from the thermodynamically available energy in the interior of planets.
It is remarkable that the law infers the magnetic fields of objects reasonably well 
from solar system planets (e.g., Earth and Jupiter) up to rapidly rotating stars such as CTTSs.
More recently, it has been applied to extrasolar planets \citep{2010A&A...522A..13R,2017ApJ...849L..12Y}.

The law is written as mathematically
\begin{equation}
\label{eq:scale_law2}
\frac{\langle B \rangle^2}{8 \pi} = c f_{\rm ohm} \langle \rho \rangle^{1/3} (F q)^{2/3},
\end{equation}
where $\langle B \rangle$ is the mean magnetic field on the dynamo surface, $c$ is a constant of proportionality, 
$f_{\rm ohm} \simeq 1 $ is the ratio of ohmic dissipation to total dissipation, 
$\langle \rho \rangle$ is the mean bulk density of a planet where the field is generated, 
$F=0.35$ is the efficiency factor of converting thermal energy to magnetic energy,
and $q= \sigma_{\rm SB}T^4_{p,e}$.
Following \cite{2009Natur.457..167C}, we adopt the typical values of Jupiter ($B_{ps}=10$ G and $q=5.4 \times 10^3$ erg s$^{-1}$ cm$^{-2}$)
and the assumption that $\langle B \rangle/B_{ps} \simeq 7$, which leads to the value of $c \simeq1.1$.

As a result, the characteristic values of $B_{ps}$ for planets around PDS 70 may be estimated as
\begin{equation}
\label{eq:Bps_th2}
1.3 \times 10^{2} \mbox{ G} \la B_{ps} \la 5.0 \times 10^2 \mbox{ G},
\end{equation}
where the upper and lower values are obtained from the consideration that 
the largest uncertainty comes from  $\langle B \rangle/B_{ps}$ with a possible range of $4 \la \langle B \rangle/B_{ps} \la 15$ \citep{2009Natur.457..167C}.

The above values are about one order of magnitude larger than the one required for magnetospheric accretion (see equation (\ref{eq:Bps_th1})).
The corresponding truncation radius of the disk ($R_{\rm T}$) for PDS 70 b/c is written as (equation (\ref{eq:R_T1}))
\begin{eqnarray}
\label{eq:R_T}
R_{\rm T} & \simeq &  5.9 R_{\rm J}   \left( \frac{ M_p}{ 10 M_{\rm J} } \right)^{-1/7}   \left( \frac{ \dot{M}_p}{ 10^{-7} M_{\rm J} \mbox{ yr}^{-1} } \right)^{-2/7}  \\ \nonumber
        & \times  &                      \left( \frac{ R_p}{ 2 R_{\rm J} } \right)^{12/7} \left( \frac{ B_{ps}}{ 130 \mbox{ G} } \right)^{4/7}.
\end{eqnarray}

It is currently unknown what strength of magnetic fields PDS 70 b/c actually have due to the lack of observations.
In addition, the direct and indirect measurements of planetary magnetic fields are very limited to date;
the measurements only for solar system planets and a very few hot Jupiters are available in the literature.
As an example, \cite{2019NatAs...3.1128C} infer that the magnetic fields of four hot Jupters range from $\sim 20$ G to $\sim 120$ G.
These estimates are based on chromospheric emission of exoplanet host stars, 
which is modulated due to the interaction with the magnetic fields of planets orbiting around these stars.
\cite{2019NatAs...3.1128C} conclude that such strong magnetic fields are in favor of the scaling law of equation (\ref{eq:scale_law2}).
However, more measurements are obviously needed to firmly confirm which scaling law (the one related to spin periods vs equation (\ref{eq:scale_law2})) would work better for accreting giant planets.

In summary, the current, empirical estimate of the magnetic field strength of the planets around PDS 70 is in the range:
\begin{equation}
\label{eq:B_ps_crit1}
20 \mbox{ G} \la B_{ps} \la 5.0 \times 10^2 \mbox{ G}.
\end{equation}

\subsection{Thermal structure of circumplanetary disks} 

In addition to planetary magnetic fields, the inner part of circumplanetary disks should be fully ionized for magnetospheric accretion;
otherwise, efficient coupling between planetary magnetic fields and the disk gas is not possible.
Here, we compute the thermal structure of circumplanetary disks around accreting, magnetized planets,
in order to further constrain the strength of planetary magnetic fields,
which is required for magnetospheric accretion.

According to Section \ref{sec:energy}, 
the temperature of the circumplanetary disk at the midplane ($T_{d,\rm mid}$) around an accreting planet can be computed as
\begin{equation}
\label{eq:disk_temp}
T^4_{d,\rm mid} \simeq T^4_{\rm vis} +T^4_{\rm irr},
\end{equation}
where viscous heating and planetary irradiation determine the disk temperature as
\begin{equation}
\label{eq:Tvis}
 T_{\rm vis}^4 = \frac{27 \kappa}{128 \sigma_{\rm SB}}     \left[  \frac{\dot{M}_p}{ 3 \pi } \left(1 - \left( \frac{r_{\rm in}}{r} \right)^{1/2} \right)  \right]^2 \frac{\Omega^3}{\alpha c_{\rm s}^2},
\end{equation}
\begin{equation}
T_{\rm irr}^4 =  \left( \frac{1}{3 \pi} \right) \left( \frac{R_p}{r} \right)^{3} T_{p,e}^4,
\end{equation}
respectively, where $\kappa$, $\alpha$, $c_{\rm s}$, and $r_{\rm in}$  are the opacity, viscosity, sound speed, and inner edge of the disk, respectively,
and $\Omega = \sqrt{GM_p/r^3}$ is the angular frequency around the planet.
In equation (\ref{eq:Tvis}), the steady state disk accretion is assumed, 
and the $\alpha-$prescription is used to characterize the disk viscosity \citep{1973A&A....24..337S}.
Also, the flat disk model is adopted for planetary irradiation,
which is generally valid for the inner part of the disk \citep{1997ApJ...490..368C}.
As shown below, this heating is not important at the inner edge of the disk for the case of PDS 70 b/c.

Equation (\ref{eq:Tvis}) contains the value of $\alpha$.
In order to compute $T_{\rm vis}$ self-consistently, 
we use the results of ideal MHD simulations \citep{2016MNRAS.457..857S}:
\begin{equation}
\label{eq:alpha}
\alpha = 11 \beta^{-0.53},
\end{equation}
where $\beta = \rho_d c_s^2/(B_p^2/8\pi)$ is the plasma beta, $\rho_d= \Sigma \Omega / (\sqrt{2 \pi} c_s)$ is the gas volume density,
and $\Sigma$ is the gas surface density of the disk. 
Note that the formula is derived from a number of ideal MHD simulations \citep{2016MNRAS.457..857S}.
Assuming that the disk viscosity in the inner edge region is driven predominantly by the dipole magnetic field of the planet (see equation (\ref{eq:B_p})),
one can compute $T_{\rm vis}$ self-consistently, using equations (\ref{eq:Tvis}) and (\ref{eq:alpha}) for given values of $B_{ps}$ and $\kappa$.

\begin{figure*}
\begin{minipage}{17cm}
\begin{center}
\includegraphics[width=8.3cm]{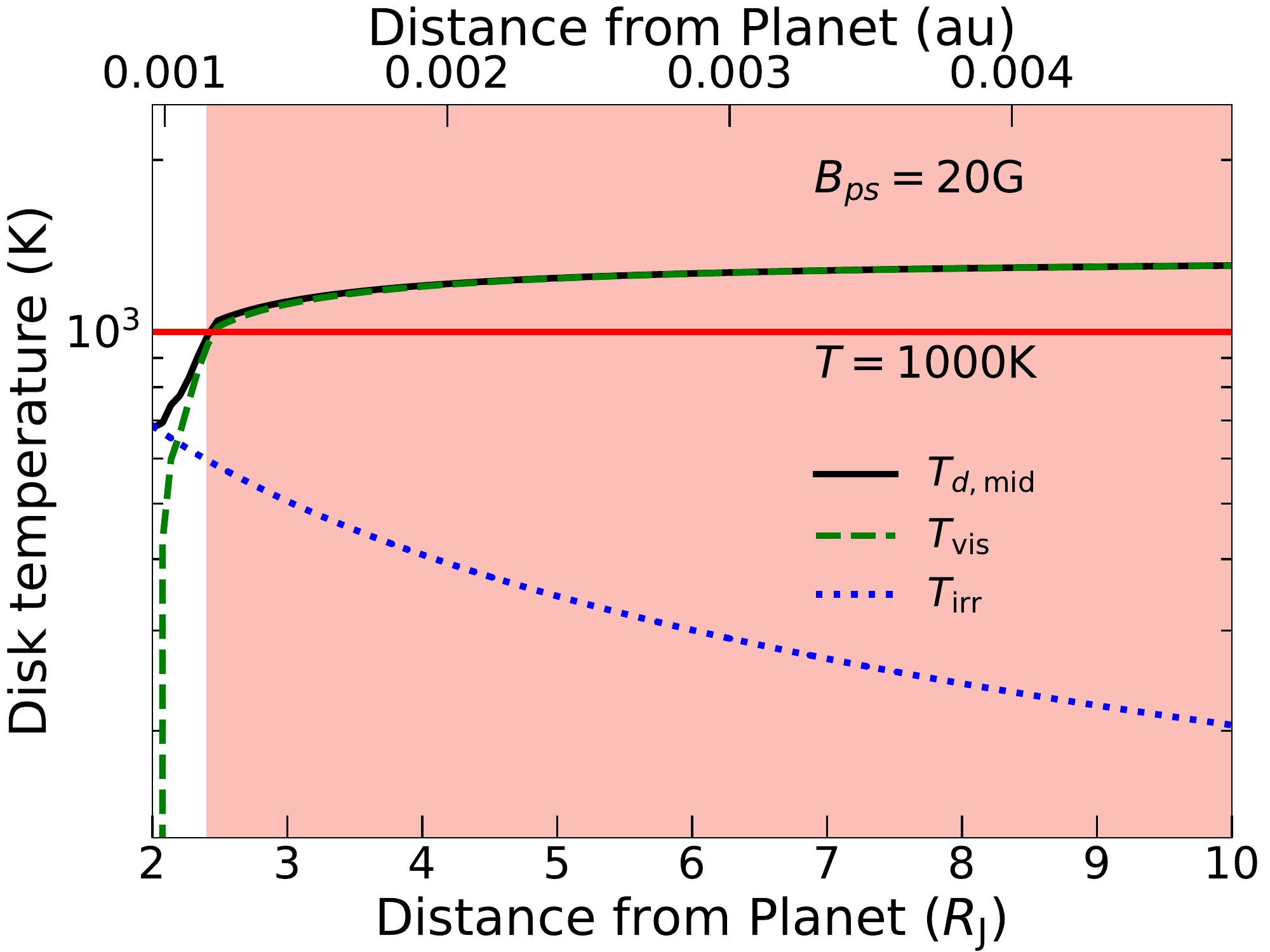}
\includegraphics[width=8.3cm]{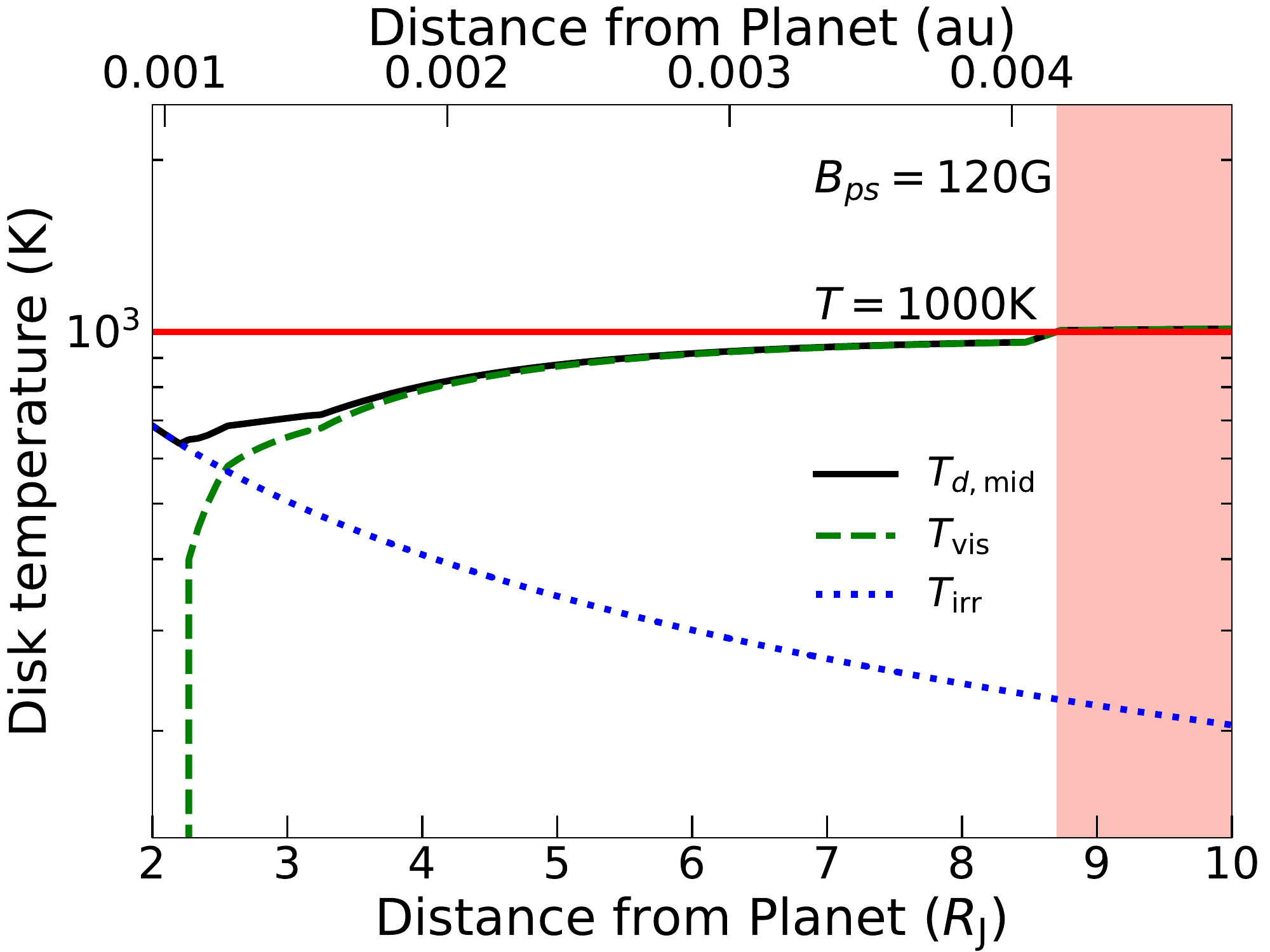}
\caption{The disk temperature at the midplane for a given value of $B_{ps}$.
The planetary magnetic fields of $B_{ps}= 20$ G and $B_{ps}= 120$ G are adopted in the left and right panels, respectively.
It is important that the disk temperature at certain disk radii becomes high enough for thermal ionization to become possible (see the red shaded region).}
\label{fig1}
\end{center}
\end{minipage}
\end{figure*}

Figure \ref{fig1} shows the resulting temperature profiles. 
In the plot, we have adopted that $r_{\rm in}=R_p$ for simplicity, which yields the value of $\eta=1/2$ in equation (\ref{eq:L_disk}).
We have also used the opacities listed in \citet{1997ApJ...486..372B};
given that the inner edge region of the disk is explored,
the regime of metal grain evaporation (i.e., $n=8$ in the paper) is most critical. 
We here consider two cases of $B_{ps}$: $B_{ps} = 20$ G and  $B_{ps} = 120$ G on the left and right panels, respectively.
Our results show that the disk temperatures at $r \ga 2.4R_{\rm J}$ and $r \ga 8.7 R_{\rm J}$ reach a threshold value ($\simeq 1000$ K) for the former and latter cases, respectively.
We have chosen $\simeq 1000$ K as the threshold value,
since at the temperature, Potassium is ionized so that thermal ionization becomes possible \citep{2014MNRAS.440...89K,2018AJ....155..178B}.
We find that as planetary magnetic fields increase, the thermally ionized region moves away from the planet.
This occurs because strong magnetic fields lead to efficient angular momentum transport of the disk (equation (\ref{eq:alpha})).
As a result, the gas surface density becomes low enough that the viscous heating eventually becomes ineffective (equation (\ref{eq:Tvis})).
This indicates that efficient coupling between planetary magnetic fields and the disk gas is achieved only in the red shaded region,
where the assumption of including viscous heating is fully justified.

As described in Section \ref{sec:mag}, magnetospheric accretion truncates the inner disk due to the magnetic pressure.
Hence, it is important to compare the truncation radius ($R_{\rm T}$) with the disk radius where $T_{d, \rm mid} \ga 1000K$.
We compute these two radii ($R_{\rm T}$ and $r(T_{d, \rm mid} = 1000$K)) as a function of $B_{ps}$.

\begin{figure}
\begin{center}
\includegraphics[width=8.3cm]{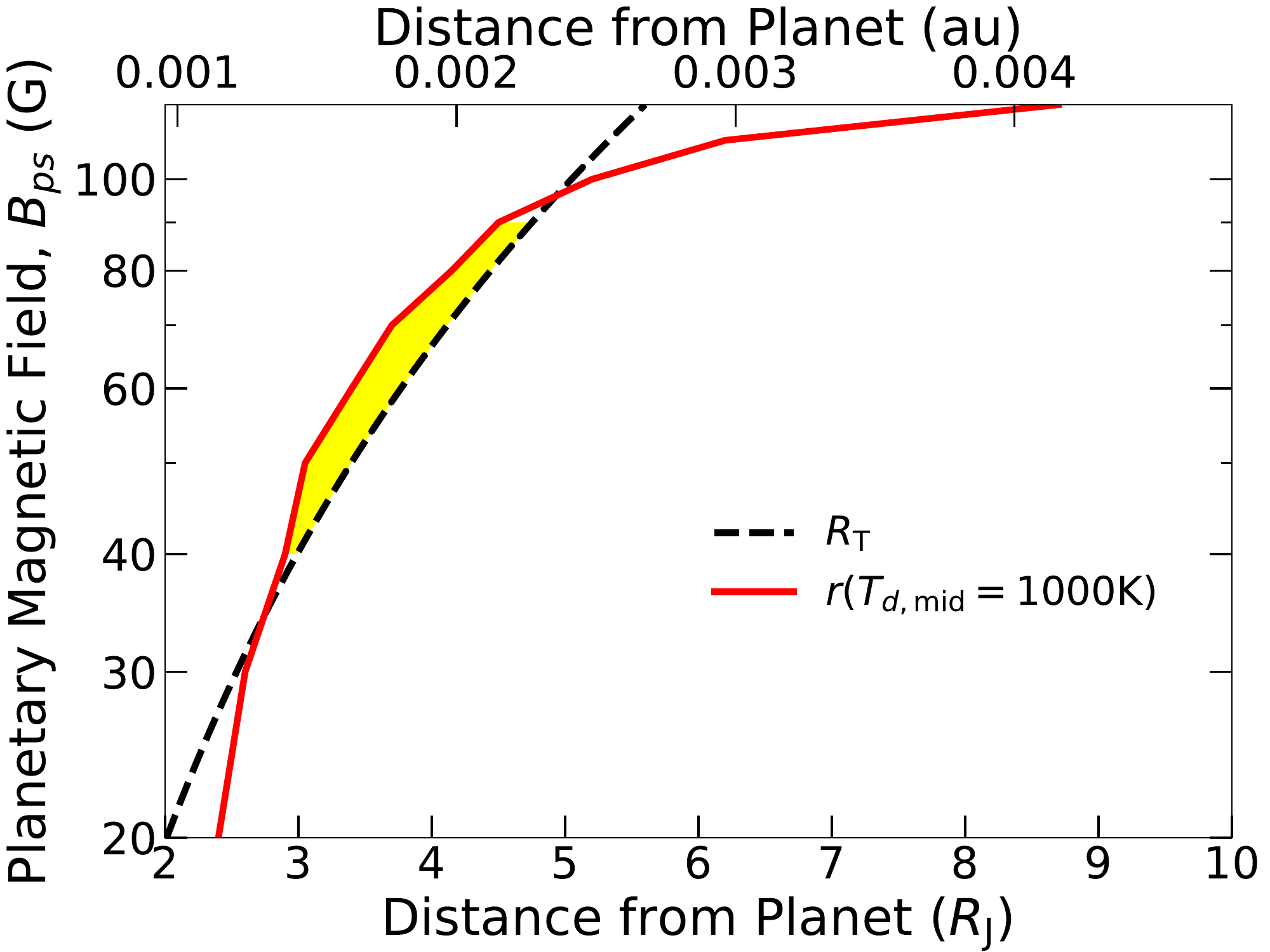}
\caption{Magnetic fields at the planetary surface ($B_{ps}$) as functions of both $R_{\rm T}$ and $r(T_{d, \rm mid} = 1000$K).
When $R_{\rm T} > r(T_{d, \rm mid} = 1000$K), the inner edge region of the disk is fully, thermally ionized,
and hence magnetospheric accretion becomes viable (see the yellow shaded region).
}
\label{fig2}
\end{center}
\end{figure}

Figure \ref{fig2} shows the results.
We find that reliable determination of $r(T_{d, \rm mid} = 1000$K) needs a proper treatment of a couple of the neighboring opacity regimes,
which makes it very inefficient to find out solutions analytically.
To circumvent the problem, we obtain the solutions by graphically searching for the interactions between equation (\ref{eq:Tvis}) and $T_{\rm vis}=1000K$ manually.
Accordingly, the behavior of $r(T_{d, \rm mid} = 1000$K) appears piece-wise slightly in Figure \ref{fig2}.
Our results show that $R_{\rm T} \ga r(T_{d, \rm mid} = 1000$K) when planetary magnetic fields are in the range of $40 \mbox{ G} \la B_{ps} \la 100 \mbox{ G}$ (see the yellow shaded region);
otherwise,  $R_{\rm T} < r(T_{d, \rm mid} = 1000$K).
Given that good coupling between planetary magnetic fields and the disk gas is necessary for magnetospheric accretion,
accreting planets should have the surface magnetic field with the range: 
\begin{equation}
\label{eq:B_ps_crit}
40 \mbox{ G} \la B_{ps} \la 100 \mbox{ G},
\end{equation}
which is narrower than the empirical estimate (see equation (\ref{eq:B_ps_crit1})).

For the planets around PDS 70, they should have the above magnetic fields,
in order to produce H$\alpha$ emission via magnetospheric accretion.

In the following section, we examine how the above range of $B_{ps}$ is plausible for the case of PDS 70 b/c, by taking into account the spin evolution of accreting, magnetized planets.

\subsection{Spin of planets} \label{sec:spin}

The spin rate of giant planets is one important quantity for exploring how giant planets accrete the gas from circumstellar and/or circumplanetary disks at their final formation stages.
Two mechanisms (disk locking and stellar winds) are often investigated to understand the spin evolution of CTTSs \citep{1991ApJ...370L..39K,2005ApJ...632L.135M}.
Since the contribution of stellar winds should be very weak for brown dwarfs and planetary-mass companions \citep{2014prpl.conf..433B},
we only examine disk locking in this work.

Disk locking occurs when the magnetic fields of young stars thread their circumstellar disks
and Stars' angular momentum is transferred to their disks \citep{1991ApJ...370L..39K}.
It is currently unconfirmed that this braking would work for magnetized planets.
However, the recent observations show that young planetary mass companions have spin rates which are a factor of few smaller than their break-up limit \citep{2018NatAs...2..138B}.
This invokes the application of disk locking to accreting young giant planets \citep{2018AJ....155..178B,2020MNRAS.491L..34G}.
Following these previous studies, we compute the spin rate of giant planets 
and determine what magnetic field (low-end vs high-end in equation (\ref{eq:B_ps_crit})) would be more reasonable for PDS 70 b/c.

The spin-up torque ($\Gamma^{\rm up}$) acting on a planet that accretes the mass from the surrounding circumplanetary disk can be written as
\begin{equation}
\Gamma^{\rm up}_p = \Gamma_{\rm acc} + \Gamma_{\rm con},
\end{equation}
where
\begin{eqnarray}
\Gamma_{\rm acc}  & = &  \dot{M}_p \sqrt{GM_p r_{\rm in}} - \dot{M}_p \sqrt{GM_p R_p}  \\ \nonumber
                                & = &  \Gamma_{\rm acc,0} \left[ 1 - \left( \frac{R_p}{r_{\rm in}}  \right)^{1/2} \right],
\end{eqnarray}
\begin{eqnarray}
\label{eq:contraction}
\Gamma_{\rm con} & = &           \frac{M_p}{2} \sqrt{\frac{GM_p} {R_p}} \frac{dR_p}{dt}                                                                                       \\ \nonumber
                               & \simeq  & \frac{\dot{M}_p}{2}  \sqrt{GM_p R_p} = \frac{1}{2} \left( \frac{R_p}{r_{\rm in}} \right)^{1/2} \Gamma_{\rm acc,0},
\end{eqnarray}
and $\Gamma_{\rm acc,0} = \dot{M}_p \sqrt{GM_p r_{\rm in}}$.
The former is due to the mass accretion from the circumplanetary disk \citep{2005ApJ...632L.135M}, 
and the latter comes from envelope contraction of the planet.
Note that in equation (\ref{eq:contraction}), it is assumed that $dR_p/dt \simeq R_p/\tau_{\rm KH}$, where $\tau_{\rm KH}$ is the Kelvin-Helmholtz timescale that regulates envelope contraction.
We have also assumed that $\tau_{\rm KH} \simeq M_p/\dot{M}_p$ \citep{2020MNRAS.491L..34G}.
Given that $R_p \la r_{\rm in}$, it may be reasonable to consider that $\Gamma^{\rm up}_{p} \simeq \Gamma_{\rm acc,0}$.

When the planet has magnetic fields that thread the surrounding circumplanetary disk,
the braking torque ($\Gamma^{\rm down}_p$) due to disk locking ($\Gamma_{\rm mag}$) may be given as \citep{1973ApJ...179..585D}
\begin{equation}
\Gamma^{\rm down}_p = \Gamma_{\rm mag} \equiv B_p^2(r_{\rm T}) r_{\rm T}^3 = B_{ps}^2 R_p^3 \left( \frac{\omega_{\rm T}}{\omega_{\rm b}} \right)^2,
\end{equation}
where $\omega_{\rm T}=\sqrt{GM_p/R_{\rm T}^3}$, and $\omega_{\rm b}=\sqrt{GM_p/R_p^3}$ is the break-up angular frequency.
A more elaborate calculation is found in \citet{2018AJ....155..178B},
where $\Gamma_{\rm mag}$ is explicitly computed, following the approach of \citet{1992MNRAS.259P..23L}.
In the approach, the azimuthal component of planetary magnetic fields is estimated,
by considering how the component builds up as the plant spins and hence the vertical component of its magnetic field is wound up.

The spin rate of the planet ($\omega_p$) can be estimated, 
by equating the above two equations ($\Gamma^{\rm up}_p=\Gamma^{\rm down}_p$)
and assuming that $\omega _{\rm T} \simeq \omega_c \equiv  \omega_p$,
where $\omega_{\rm c}=\sqrt{GM_p/r_{\rm c}^3}$, and $r_{\rm c}$ is the corotational radius of the planet.
Then, $\omega_p$ is given as
\begin{eqnarray}
\label{eq:spin}
\omega_{p} & \simeq          & \omega_{\rm b}  \left[ \frac{1}{2} \left( \frac{r_{\rm in}}{R_p} \right) \left( \frac{R_{\rm T}}{R_p} \right)^{-7}  \right]^{1/4} \\ \nonumber
                         & \simeq & 3.0 \times 10^{-1} \omega_{\rm b}  \left( \frac{R_{\rm T}/R_p}{2} \right)^{-3/2},  
\end{eqnarray}
where $r_{\rm in} \simeq R_{\rm T} \simeq 2 R_p$ has been adopted (see Figure \ref{fig2}).
Note that our estimate is very  approximate, since the magnetic torque ($ \Gamma_{\rm mag}$) vanishes at the corotational radius.

As \cite{2018AJ....155..178B} pointed out, the budget of angular momentum of circumplanetary disks is much smaller than that of planets at the break-up limit.
This implies that the angular momentum transferred from the planets to the circumplanetary disks should be further transferred to the circumstellar disks.
Also, it is well known that magnetic braking is so efficient that planets achieve a steady state for spin evolution very quickly \citep{1991ApJ...370L..39K,2018AJ....155..178B}.

Equation (\ref{eq:spin}) indicates that if disk locking is at play for accreting giant planets, 
the spin rate of these planets becomes about 20 \% of the break-up limit for the case that $B_{ps} \simeq 100$ G (that is, $R_{\rm T} \simeq 2.5 R_p$)
and about 46 \% of the break-up limit for the case that $B_{ps} \simeq 40$ G (that is, $R_{\rm T} \simeq 1.5 R_p$).
When the mass accretion onto the planets ceases and their envelopes shrink further from $2 R_{\rm J}$ to $1R_{\rm J}$, 
the spin rate can be increased by a factor of 4, under the assumption that the planet mass and moment of inertia are constant.
Note that the subsequent envelope contraction occurs on a very long ($\sim 100$ Myr) timescale \citep{2012ApJ...745..174S}.
Taking into account the subsequent evolution,
higher magnetic fields ($\sim 100$ G) are more preferred to broadly reproduce the observational results of \citet{2018NatAs...2..138B}.

One may consider that the above estimates should be viewed as an upper limit;
as disk evolution proceeds, the accretion rate decreases, which in turn increases $R_{\rm T}$ (see equation (\ref{eq:R_T})).
This slows down $\omega_p$ as long as good coupling between planetary magnetic fields and the disk gas is established.
This consideration is sensible.
However, we find that the condition of the good coupling cannot be met if the accretion rate becomes lower than a certain value;
for the case that $B_{ps} = 40$ G, $R_{\rm T} < r(T_{d, \rm mid}= 1000$ K) when $\dot{M}_p = 7 \times 10^{-8}$ M$_{\rm J}$ yr$^{-1}$ or lower,
and for the case that $B_{ps} = 100$ G, $R_{\rm T} < r(T_{d, \rm mid} =1000$ K) when $\dot{M}_p = 9 \times 10^{-8}$ M$_{\rm J}$ yr$^{-1}$ or lower (also see Figure \ref{fig2}).
In other words, our estimates provide an upper limit, but should be reasonable for inferring the spin rate of matured planets.

Thus, our calculation suggests that strong magnetic fields ($\sim 100$ G) are more plausible for PDS 70 b/c, 
and leads to the following conclusion:
while disk locking leads to spin-down of giant planets during the process of forming,\footnote{
Even when the planet radius is larger than $2 R_{\rm J}$, which occurs at the early stages of giant planet formation,
disk-locking can decrease the spin rate down to a few \% of the break-up limit,
if the inner edge region of circumplanetary disks is fully ionized \citep{2020MNRAS.491L..34G}.}
the subsequent envelope contraction increases their spin rates as with the case for brown dwarfs \citep{2018ApJ...859..153S}.

\section{Circumplanetary disks around magnetized planets} \label{sec:disk}

We have confirmed above that magnetospheric accretion is possible for PDS 70 b/c, 
and strong ($\sim 10^2$ G) planetary magnetic fields are preferred to better understand the recent observations.
In this section, we further explore the properties of circumplanetary disks around such magnetized planets.
We especially focus on the inner edge region ($5 R_{\rm J} \la r \la 10 R_{\rm J}$) of the disks,
where good coupling between planetary magnetic fields and the disk gas is surely achieved due to thermal ionization.
Given that the magnetic field profile of accreting planets may deviate considerably from the dipole one in circumplanetary disks,
we consider two cases, which are discussed below. 

\subsection{Gas accretion flow around planets} \label{sec:disk_flow}

We first consider gas accretion flow around planets.
It is currently inconclusive how planets accrete the gas from circumstellar and/or circumplanetary disks.
However, many 3D hydrodynamical simulations show that gas accretion flow onto circumplanetary disks comes from circumstellar disks in the vertical direction 
\citep[e.g.,][]{2012ApJ...747...47T,2014ApJ...782...65S}.
The flow generates shock at the surface of the circumplanetary disks and moves inward.
The inward movement occurs because the shocked gas has specific angular momentum that is smaller than that of the local Keplerian motion.
On the other hand, the gas in the midplane region either is captured in the horse-shoe orbit around the planets, or spirals outward and eventually escapes from the Hill radius of the planets.
This suggests that circumplanetary disks may experience layered accretion.

We here adopt the above picture and compute the mass flux onto circumplanetary disks originating from circumstellar disks ($\dot{M}_{p}^{\rm CSD}$).
To proceed, we adopt the approach of \cite{2016ApJ...823...48T}, where the results of two different hydrodynamical simulations are combined;
the one computes the accretion rate onto the system of a planet and its circumplanetary disk from the parental circumstellar disk \citep{2002ApJ...580..506T},
and the other calculates the reduction in the surface density of the circumstellar disk due to disk-planet interaction \citep{2015ApJ...806L..15K}.
Then, $\dot{M}_{p}^{\rm CSD}$ is written as 
\begin{eqnarray}
\label{eq:Mpdot_CSD}
\dot{M}_{p}^{\rm CSD} & =          & \frac{8.5}{3 \pi} \left( \frac{c_s^{\rm CSD}}{v_{\rm Kep}^{\rm CSD}} \right) \left( \frac{M_p}{M_s} \right)^{-2/3}  \dot{M}_{s} \\ \nonumber
                                 & \simeq & 1.8 \times 10^{-7}  M_{\rm J} \mbox{ yr}^{-1} \left( \frac{c_s^{\rm CSD} / v_{\rm Kep}^{\rm CSD}}{8.9 \times 10^{-2}} \right)   \\ \nonumber
                                 & \times  &   \left( \frac{M_p}{10M_{\rm J}} \right)^{-2/3} \left( \frac{\dot{M}_{s}}{1.4 \times 10^{-7} M_{\rm J} \mbox{ yr}^{-1}}  \right) ,
\end{eqnarray}
where $c_s^{\rm CSD}$ and $v_{\rm Kep}^{\rm CSD}$ are the sound speed and the Keplerian velocity of the circumstellar disk gas at the position of the planet, 
and $M_s$ and $\dot{M}_s$ are the mass of the central star and the disk accretion rate onto the star, respectively.
We have adopted that $c_s^{\rm CSD}/v_{s, \rm Kep}^{\rm Kep}=8.9 \times 10^{-2}$ and $M_s=0.76~ M_{\odot}$, 
following \citet{2018AA...617A..44K} which examine the properties of the circumstallar disk around PDS 70.
The value of $\dot{M}_{s}$ is taken from \citet{2020ApJ...892...81T} 
which suggest that $\dot{M}_{s}$ of PDS 70 lies within the range of $0.6-2.2 \times 10^{-7} M_{\rm J} \mbox{ yr}^{-1}$.
Note that while \cite{2014Icar..232..266M} derive a different formula of $\dot{M}_{p}^{\rm CSD}$, by explicitly considering the geometry of gas accretion flow,
\cite{2019ApJ...876L..32H} demonstrate that the resulting accretion rate becomes comparable to the one computed from equation (\ref{eq:Mpdot_CSD}) in the Jovian-mass regime.

It is important that $\dot{M}_{p}^{\rm CSD} \simeq \dot{M}_p$ within the range of $\dot{M}_{s}$.
This suggests that the gas originating from the circumstellar disk may be accreted onto the planet through the circumplanetary disk;
equivalently, it may not be unreasonable to anticipate that the steady state accretion assumption broadly holds for circumplanetary disks around PDS 70 b/c.
It should be noted that this is the first attempt of applying equation (\ref{eq:Mpdot_CSD}) to the observed system,
and hence we should consider that this steady state assumption is verified only for the current epoch of giant planet formation for this particular target;
the validity of equation (\ref{eq:Mpdot_CSD}) is confirmed, using the results of numerical simulations, 
where the accretion rate of circumstallr disks is about three orders of magnitude higher than that adopted in this work \citep[see their figure 1]{2016ApJ...823...48T}.
However, it remains to be confirmed whether the steady state accretion assumption is reasonable for other systems.

In the following sections, we adopt the assumption and compute the gas properties of the circumplanetary disks at the inner edge region.

\subsection{Dipole magnetic field case} \label{sec:dipole_b-field}

We here consider the case that planetary magnetic fields in the disk inner edge region can still be approximated as dipole, that is, $B_p \propto r^{-3}$.
As described above, it is reasonable to expect that planetary magnetic fields deviate considerably from the dipole profile due to the interaction with the gas in circumplanetary disks
as with the case for CTTS.
However, the actual field profile is unknown, even for CTTS.
Recently, \citet{2019A&A...629L...1H} attempt to constrain the profile from the population of close-in giant planets,
and find that stellar magnetic fields in the inner edge region of the circumstallar disks may be bracketed by two characteristic profiles:
one is the dipole one, and the other is $B_p \propto r^{-2}$.
The latter corresponds to a steady state solution, 
where advection and diffusion of magnetic flux in the radial direction compete with each other.
Motivated by the work, we consider the former and the latter in this and next sections, respectively.

As described in Section \ref{sec:disk_flow}, the steady state disk accretion model is applicable to the circumplanetary disks around PDS 70 b/c,
and it is reasonable to assume the ideal MHD limit (Figure \ref{fig2}). 
Then, the disk accretion rate is given as
\begin{equation}
\label{eq:Mdot}
\dot{M}_p = 3 \pi \nu \Sigma \left[ 1 - \left( \frac{r}{5 R_{\rm J}} \right)^{-1/2} \right]^{-1},
\end{equation}
where $\nu=\alpha c_s^2/\Omega$ is the effective viscosity, and it is adopted that $r_{\rm in}= 5 R_{\rm J}$; equivalently, $B_{ps}=100$ G, based on Figure \ref{fig2}.
The value of $\alpha$ is computed from equation (\ref{eq:alpha}).

Assuming that $T_{d,\rm mid} \simeq 1000$ K in the disk midplane,
the disk properties at the inner edge region are given as 
\begin{eqnarray}
\beta & \simeq & 6.5 \times 10^3   \left( \frac{ M_p}{ 10 M_{\rm J} } \right)^{2.1}   \left( \frac{ \dot{M}_p}{ 10^{-7} M_{\rm J} \mbox{ yr}^{-1} } \right)^{2.1}  \\ \nonumber
                & \times  &                      \left( \frac{ R_p}{ 2 R_{\rm J} } \right)^{-12.8}  \left( \frac{ T_{d,e} }{ 1000 \mbox{ K}} \right)^{-1} \left( \frac{ B_{ps}}{ 100 \mbox{ G} } \right)^{-4.3}
                                                        \left( \frac{ r}{ 10 R_{\rm J} } \right)^{6.4}, 
\end{eqnarray}
\begin{eqnarray}
\alpha & \simeq & 1.0 \times 10^{-1}   \left( \frac{ M_p}{ 10 M_{\rm J} } \right)^{-1.1}   \left( \frac{ \dot{M}_p}{ 10^{-7} M_{\rm J} \mbox{ yr}^{-1} } \right)^{-1.1}  \\ \nonumber
                & \times  &                      \left( \frac{ R_p}{ 2 R_{\rm J} } \right)^{6.8}  \left( \frac{ T_{d,e} }{ 1000 \mbox{ K}} \right)^{0.6} \left( \frac{ B_{ps}}{ 100 \mbox{ G} } \right)^{2.3}
                                                        \left( \frac{ r}{ 10 R_{\rm J}  } \right)^{-3.4}, 
\end{eqnarray}
\begin{eqnarray}
\label{eq:sigma1}
\Sigma & \simeq & 1.5 \mbox{ g cm}^{-2}    \left( \frac{ M_p}{ 10 M_{\rm J} } \right)^{1.6}   \left( \frac{ \dot{M}_p}{ 10^{-7} M_{\rm J} \mbox{ yr}^{-1} } \right)^{2.1}  \\ \nonumber
                & \times  &                      \left( \frac{ R_p}{ 2 R_{\rm J} } \right)^{-6.8}  \left( \frac{ T_{d,e} }{ 1000 \mbox{ K}} \right)^{-1.6} \left( \frac{ B_{ps}}{ 100 \mbox{ G} } \right)^{-2.3}
                                                        \left( \frac{ r}{ 10 R_{\rm J} } \right)^{1.9}, 
\end{eqnarray}
\begin{eqnarray}
n_{d,H} & \simeq & 1.9 \times 10^{13} \mbox{ cm}^{-3}    \left( \frac{ M_p}{ 10 M_{\rm J} } \right)^{2.1}   \left( \frac{ \dot{M}_p}{ 10^{-7} M_{\rm J} \mbox{ yr}^{-1} } \right)^{2.1}  \\ \nonumber
                & \times  &                      \left( \frac{ R_p}{ 2 R_{\rm J} } \right)^{-6.8}  \left( \frac{ T_{d,e} }{ 1000 \mbox{ K}} \right)^{-2.1} \left( \frac{ B_{ps}}{ 100 \mbox{ G} } \right)^{-2.3}
                                                        \left( \frac{ r}{10 R_{\rm J} } \right)^{0.4},
\end{eqnarray}
where the disk radius ($r$) is normalized by $r = 10 R_{\rm J}$, $n_{d,H}= \rho_d/ \mu$ is the number density of hydrogen, and $\mu$ is the mean molecular weight.
Note that the usage of equation (\ref{eq:alpha}) is justified because the formula is obtained for the range of $10 \le \beta \le 10^5$ \citep{2016MNRAS.457..857S};
equivalently, the ideal MHD assumption is justified in the inner edge region. 

It is worth mentioning that $\Sigma$ becomes an increasing function of $r$ in the region, where planetary magnetic fields play an important role in disk accretion.
This occurs because the value of $B_p$ and hence $\alpha$ decrease as the distance from the planet increases.
The positive slope of $\Sigma$ is also expected in the inner edge region of circumstellar disks around CTTSs due to stellar dipole fields \citep[e.g.,][]{2019A&A...629L...1H}.

We now compare the above disk properties with other disk models.
In the literature, three kinds of models are currently available; 
the first kind is the so-called minimum mass {\it sub}nebula model \citep[MMSN,][]{1982Icar...52...14L}.
This model is the counterpart of the minimum mass {\it solar} nebula model \citep{1981PThPS..70...35H}
and is derived from the mass and orbital distributions of the four Galilean moons.
The second kind of the model is the gas-starved model \citep[e.g.,][]{2002AJ....124.3404C}
and is proposed to resolve the issues of the MMSN model;
the MMSN model has difficulty in reproducing the composition of the Galilean moons
due to high disk temperatures as a result of high gas surface densities.
In the gas-starved model, such difficulty is resolved by continousely lowering the surface density and temperature of circumplanetary disks over the disk lifetime,
such that H$_2$O ice that is the building block of the moons, can be present at the current regular satellite region.
The last kind of the model are numerical simulations \citep[e.g.,][]{2017ApJ...842..103S}.
It is obvious that numerical simulations provide the most detailed properties of disks.
However, the validity of the employed assumptions currently cannot be examined due to the lack of disk observations;
PDS 70 c is the only target that robustly exhibits the presence of the circumplanetary disk \citep{2019ApJ...879L..25I,2021ApJ...916L...2B}.
Therefore, we here focus on the empirically derived models, that is, the MMSN and gas-starved models.

These two models predict that $\Sigma(r = 10 R_{\rm J}) \simeq 1.0 \times 10^6$ g cm$^{-2}$ for the MMSN model \citep{1996Icar..123..404T} 
and $\Sigma(r = 10 R_{\rm J}) \la 3 \times 10^2$ g cm$^{-2}$ for the gas-starved model \citep{2002AJ....124.3404C}.
It is important that our estimates are much lower than these values;
this may arise partly because magnetic fields of planets around PDS 70 trigger efficient angular momentum transport in their circumplanetary disks,
and partly because the low accretion rate of the circumstallar disk around PDS 70 eventually reduces the surface density of the circumplanetary disks around PDS 70 b/c.

\subsection{Steady state solution case} \label{sec:steady_b_field}

We here consider the steady state solution case for planetary magnetic fields, that is, $B_p \propto r^{-2}$.
We adopt the same parameters and assumptions as in Section \ref{sec:dipole_b-field}.
In order to reliably compare this case with the dipole case,
we impose conservation of the total magnetic flux that threads the disk inner edge region ($5 R_{\rm J} \la r 10 R_{\rm J}$).
As a result, $B_p$ at $r= 5 R_{\rm J}$ for this case is about 0.7 times weaker than that for the dipole case.

The resulting disk properties at the inner edge region are written as
\begin{eqnarray}
\beta & \simeq & 8.2 \times 10^1   \left( \frac{ M_p}{ 10 M_{\rm J} } \right)^{2.1}   \left( \frac{ \dot{M}_p}{ 10^{-7} M_{\rm J} \mbox{ yr}^{-1} } \right)^{2.1}  \\ \nonumber
                & \times  &                      \left( \frac{ R_p}{ 2 R_{\rm J} } \right)^{-8.5}  \left( \frac{ T_{d,e} }{ 1000 \mbox{ K}} \right)^{-1.1} \left( \frac{ B_{ps}}{ 100 \mbox{ G} } \right)^{-4.3}
                                                        \left( \frac{ r}{ 10 R_{\rm J} } \right)^{2.1}, 
\end{eqnarray}
\begin{eqnarray}
\alpha & \simeq & 1.1   \left( \frac{ M_p}{ 10 M_{\rm J} } \right)^{-1.1}   \left( \frac{ \dot{M}_p}{ 10^{-7} M_{\rm J} \mbox{ yr}^{-1} } \right)^{-1.1}  \\ \nonumber
                & \times  &                      \left( \frac{ R_p}{ 2 R_{\rm J} } \right)^{4.5}  \left( \frac{ T_{d,e} }{ 1000 \mbox{ K}} \right)^{0.6} \left( \frac{ B_{ps}}{ 100 \mbox{ G} } \right)^{2.3}
                                                        \left( \frac{ r}{ 10 R_{\rm J}  } \right)^{-1.1}, 
\end{eqnarray}
\begin{eqnarray}
\label{eq:sigma2}
\Sigma & \simeq & 1.5 \times 10^{-1} \mbox{ g cm}^{-2}    \left( \frac{ M_p}{ 10 M_{\rm J} } \right)^{1.6}   \left( \frac{ \dot{M}_p}{ 10^{-7} M_{\rm J} \mbox{ yr}^{-1} } \right)^{2.1}  \\ \nonumber
                & \times  &                      \left( \frac{ R_p}{ 2 R_{\rm J} } \right)^{-4.5}  \left( \frac{ T_{d,e} }{ 1000 \mbox{ K}} \right)^{-1.6} \left( \frac{ B_{ps}}{ 100 \mbox{ G} } \right)^{-2.3}
                                                        \left( \frac{ r}{ 10 R_{\rm J} } \right)^{-0.4}, 
\end{eqnarray}
\begin{eqnarray}
n_{d,H} & \simeq & 1.8 \times 10^{13} \mbox{ cm}^{-3}    \left( \frac{ M_p}{ 10 M_{\rm J} } \right)^{2.1}   \left( \frac{ \dot{M}_p}{ 10^{-7} M_{\rm J} \mbox{ yr}^{-1} } \right)^{2.1}  \\ \nonumber
                & \times  &                      \left( \frac{ R_p}{ 2 R_{\rm J} } \right)^{-4.5}  \left( \frac{ T_{d,e} }{ 1000 \mbox{ K}} \right)^{-2.1} \left( \frac{ B_{ps}}{ 100 \mbox{ G} } \right)^{-2.3}
                                                        \left( \frac{ r}{10 R_{\rm J} } \right)^{-1.9}.
\end{eqnarray}
We again confirm that the usage of equation (\ref{eq:alpha}) is verified.

Our calculations show that the value of $\Sigma$ is very low;
this is caused by planetary magnetic fields as in Section \ref{sec:dipole_b-field}.
For this case, however, $\Sigma$  becomes a decreasing function of $r$.
We find that a positive slope of $\Sigma$ is achieved when $B_p \propto r^n$ with $n < -2.2$.

In the following section, we discuss the implications of these disk models for satellite formation.

\subsection{Implications for satellite formation}

We have so far focused on the properties of circumplanetary disks at the inner edge region.
Therefore, it is inappropriate to explore how satellites form in the disks {\it globally} \citep[c.f.][]{2020ApJ...894..143B}.
The inner edge region, however, is known to play an important role in the formation of close-in planets \citep[e.g.,][]{2015A&A...584L...1O,2017MNRAS.470.1750I,2021arXiv210607058A}.
Motivated by this, we here discuss implications of our findings for satellite formation.

The most unique feature of our results is that $\Sigma$ takes a very low value in the region,
where planetary magnetic fields are important.
This finding is valid as long as the host planets have magnetic fields strong enough to undergo magnetospheric accretion,
and is independent of the profile of their magnetic fields.
Given that such a low-$\Sigma$ region overlaps with the current locations of Galilean moons of Jupiter,
the immediate conclusion is derived that radial movement of satellite-forming materials is needed,
as with the case for planet formation.

Another interesting feature is that $\Sigma$ can become an increasing function of $r$ in the inner edge region of the disk.
This feature depends on the profile of planetary magnetic fields in the disk, as demonstrated in Section \ref{sec:steady_b_field}.
When magnetospheric accretion is realized, however,
the inner disk is truncated, and hence it is natural to assume that the positive slope of $\Sigma$ is achieved there.
The following discussion is developed under this assumption.

We first point out that with some exceptions \citep[e.g.,][]{1996Icar..123..404T}, 
many of the currently existing models do not exhibit such structures \citep[e.g.,][]{2014ApJ...788..129I,2017ApJ...842..103S},
simply because the effect of planetary magnetic fields is not considered in detail.

We then discuss its effect on migration of (proto)satellites;
it is well known that disk-planet interaction and the resulting migration are very sensitive to the surface density (and disk temperature) profiles 
\citep[e.g.,][]{2010MNRAS.401.1950P,2011MNRAS.417.1236H}.
When the surface density has a positive radial gradient, the migration can be reversed due to the corotational torque \citep{2006ApJ...642..478M}.
It is interesting that the current position of Io is about $6 R_{\rm J}$,
which coincides with the truncation radius of the disks around PDS 70 b/c.
This implies that proto-Jupiter might have had a similar strength ($\sim 100$ G) of magnetic fields.
Thus, the current location of Io may be indicative of the presence of traps for migrating (proto)satellites in circumplanetary disks.

The positive slope of $\Sigma$ also affects the spatial distribution of dust in circumplanetary disks, as with the case for circumstellar disks \citep[e.g.,][]{2007ApJ...664L..55K}.
The actual size and spatial distributions of dust can be explored realistically only when the gas motion of circumplanetary disks is properly taken into account \citep{2018ApJ...866..142D}.
However, the positive gradient of $\Sigma$ can produce dust traps, and the presence of the traps can affect the detectability of circumplanetary disks \citep{2018MNRAS.479.1850Z}.
Intriguingly, the pile-up of dust particles at the inner edge region may trigger the inside-out satellite formation \citep{2012Sci...338.1196C},
as suggested for close-in super-Earths \citep{2014ApJ...780...53C}.
Our results are therefore important for investigating the dust distribution in circumplanetary disks.

It should be noted that since the gas temperature at the inner disk region is higher than the sublimation temperature of icy materials (see Figure \ref{fig1}),
it is still unclear how satellites such as Europa obtained volatiles (e.g., water);
icy dust (or pebble) sized particles should sublimate in the inner disk region.
While (proto)Europa might have initially formed in the outer part of the disk and subsequently migrated to the current location,
evaporation of volatiles from it might be possible.
Thus, further investigations are needed for fully understanding satellite formation.

\section{Summary \& discussion} \label{sec:disc}

We have explored the properties of accreting, magnetized giant planets surrounded by circumplanetary disks.
This is motivated by the recent high spatial/spectral resolution observations which shed light on the final formation stages of such planets.
These observations include the H$\alpha$ detections around PDS 70, the measurements of spin rates of young giants, 
and indirect estimates of magnetic fields of hot Jupiters.

We have begun with determination of heating sources for the circumplanetary disks
and shown that both viscous heating and planetary irradiation can be equally important.
We have then computed the strength of planetary magnetic fields that is needed for magnetospheric accretion.
Our calculations find that at least an order of a few 10 G are required (see equation (\ref{eq:Bps_th1})).
It is interesting that such strength can be achieved when planets spin fast enough.
Thus, it can be concluded that giant planets may naturally experience magnetospheric accretion at their final formation stages,
where the planetary spin is accelerated, following mass growth and envelope contraction.
The prediction of rapid rotators, however, is not compatible with the observations (see equation (\ref{eq:spin})).
In summary, stronger planetary magnetic fields are desired to draw a consistent picture.

The stronger magnetic fields are anticipated, based on the scaling law of \citet{2009Natur.457..167C}.
Using the physical parameters inferred for PDS 70 b/c (see Table \ref{table1}),
we find that accreting giant planets can possess an order of a few 100 G (see equation (\ref{eq:Bps_th2}));
the scaling law can reproduce the field estimates of hot Jupiters as well.

In addition to planetary magnetic fields,
it is critical to determine whether the inner edge region of circumplanetary disks is ionized enough
that efficient coupling between the magnetic fields and the disk gas is realized.
We have self-consistently computed the disk temperature at the midplane, using the results of ideal MHD simulations,
and obtained a better constrain on the strength of planetary magnetic fields (see equation (\ref{eq:B_ps_crit}), also see Figure \ref{fig2});
the preference  is given for the high-end ($\sim 100$ G) of the range.

Such strong fields lead to a unified interpretation of all the currently available observations:
The field strength surely supports the action of magnetospheric accretion that is an origin of the observed H$\alpha$ observations,
and the field strength naturally explains the slow spin rate of young giants via disk locking and the subsequent envelope contraction.
Thus, the final stages of giant planet formation are very likely comparable to those of low-mass stars such as brown dwarfs.

We have also computed the properties of circumplanetary disks around accreting, magnetized giant planets.
We have focused only on the inner edge region ($5 R_{\rm J}  \la r \la 10 R_{\rm J}$).
Given that planetary magnetic fields may deviate from the dipole profile in the region due to the interaction with the disk gas,
we have considered two cases: one is the dipole ($\propto r^{-3}$), and the other is $\propto r^{-2}$, following a previous study.
We find that the resulting disk properties exhibit unique features which are quite different from the canonical models of circumplanetary disks:
a very low value of the gas surface density, and the surface density increases with increasing the distance from the host planet (e.g., see equation (\ref{eq:sigma1})).
This profile can lead to the proposition of both satellite and dust traps,
which are crucial for better understanding satellite formation in circumplanetary disks.
Such a profile has not been found in most of the previous studies, since they do not consider planetary magnetic fields in detail.
Our results therefore indicate that satellite formation should be highly dynamics as with the case for extrasolar and solar planetary system formation.
Note that due to the high gas temperature at the trap location, the origin of volatiles on satellites such as Europa is still unclear.

We must admit that our models are very simple, and more detailed modeling and/or simulations are needed to verify our calculations.
For instance, we have computed the disk temperature, assuming that $r_{\rm in} = R_p$ (see equation (\ref{eq:Tvis}) in Section \ref{sec:mod}).
This assumption becomes invalid when the inner disk is truncated due to planetary magnetic fields;
for this case, some of gravitational energy would be released at the boundary between the planetary magnetic fields and the inner edge.
Our model does not consider such a heating source explicitly;
instead, we have imposed the conservation of energy, by assuming that $r_{\rm in} = R_p$,
which might take into account some effect of the heating.
It is obvious that a more self-consistent approach would be demanded to accurately compute the disk temperature at the midplane.
Another simplification is that we have adopted the opacities that are derived from circumstallar disks;
there is no guarantee that the opacities are applicable to circumplanetary disks as well. 
Given that the gas and dust properties of the circumplanetary disks are unknown,
we consider that its usage is currently acceptable;
if satellite formation may take place as for planet formation, 
the properties of circumplanetary disks may be comparable to that of circumstellar disks.
Observational discoveries and characterization of circumplanetary disks are truly longed for.

Some caution is obviously needed to evaluate the results of our calculations.
We have leveraged the results of MHD simulations that are carried out for modeling circumstellar disks (Section \ref{sec:disk}).
As described above, however, the gas and dust properties of circumplanetary disks may be different from those of parental circumstellar disks;
the gas accretion flow from circumstellar disks to circumplanetary disks originates only from the high latitude,
where the disk gas is exposed to high energy photons and the dust abundance is very low due to dust growth and settling.
It is also crucial to investigate what geometry of planetary and disk magnetic fields look like, by running detailed MHD simulations.

In the near future, more observational data and detailed modeling/simulations will become available.
Our efforts made in this work will serve as a stepping stone 
not only for guiding these studies, but also for developing a better understanding of the formation of giant planets and satellites around them.

\begin{acknowledgments}

The authors thank an anonymous referee for useful comments, which significantly improve the quality of our manuscript.
This research was carried out at the Jet Propulsion Laboratory, California Institute of Technology,
under a contract with the National Aeronautics and Space Administration.
The authors thank Yuuhiko Aoyama and Jun Hashimoto for stimulating discussions.
Y.H. is supported by JPL/Caltech.
K.D.K. was supported by JSPS KAKENHI grant 19K14779.   

\end{acknowledgments}

\bibliographystyle{aasjournal}
\bibliography{adsbibliography}    



\end{document}